\documentstyle[aps,prl,twocolumn,epsfig]{revtex}
\begin{document}
\input{psfig.sty}
\draft

\newcommand{\bQ}{\mbox{${\bf Q}$}}
\newcommand{\h}{\mbox{$\frac{1}{2}$}}
\newcommand{\th}{\mbox{$\frac{3}{2}$}}
\def\bolda{\mbox{\boldmath$a$}}
\def\boldb{\mbox{\boldmath$b$}}
\def\boldc{\mbox{\boldmath$c$}}
\def\boldQ{\mbox{\boldmath$Q$}}
\def\boldq{\mbox{\boldmath$q$}}
\def\btau{\mbox{\boldmath$\tau$}}

\twocolumn[\hsize\textwidth\columnwidth\hsize\csname
@twocolumnfalse\endcsname

\title{Continuum in the spin excitation spectrum of a Haldane
chain (erratum). }

\author{I.~A.~Zaliznyak$^{1}$, S.-H.~Lee$^{2,3}$, S.~V.~Petrov$^4$}
\address{
 $^1$Department of Physics, Brookhaven National Laboratory, Upton,
 New York 11973-5000 \\
 $^2$National Institute of Standards and Technology, Gaithersburg,
 Maryland 20899\\
 $^3$Department of Physics, University of Maryland, College Park,
 Maryland 20742\\
 $^4$P.~Kapitza Institute for Physical Problems, ul. Kosygina, 2,
 Moscow, 117334 Russia
 }

\date{\today}
\maketitle



]

In Ref. \onlinecite{Zaliznyak2001} we reported inelastic neutron scattering
study of the spin excitation spectrum in the S=1 quasi-one-dimensional
antiferromagnet CsNiCl$_3$. A cross-over from the coherent, single-particle
magnon excitation at wavevectors close to $q = \pi$, to a continuum dominated by
the two-particle states at small wavevectors, $q\lesssim 0.5\pi$, was clearly
demonstrated by the measured intensities shown in the Fig. 2 of that paper.
For the sake of comparison, the two-magnon lineshapes predicted from the
nonlinear $\sigma$-model by Affleck and Weston in Ref.
\onlinecite{AffleckWeston} were also shown, by the dotted lines, in the figure.

\begin{figure}[b] \noindent
\parbox[b]{3.4in}{\psfig{file=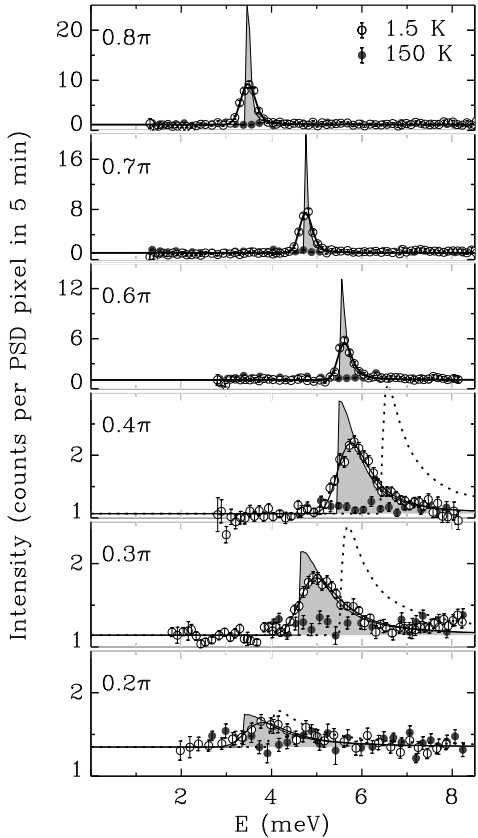,height=4.1in}
\vspace{0.25in} \noindent%
\caption{Corrected version of the Fig. 2 in Ref.
\protect\cite{Zaliznyak2001}. Dotted lines which show the prediction of Ref.
\protect\cite{AffleckWeston} were corrected. }}
\end{figure} \vspace{-0.1in}\noindent

It recently came to our attention that theoretical curves shown in Fig. 2 of our
paper \cite{Zaliznyak2001} are slightly incorrect. This is clear if one compares
the onset of the dotted line in Fig. 2 with the 2-magnon boundary shown in Fig.
3 of the paper, which is at lower energy. The error in Fig. 2 came about because
an incorrect value of the Haldane gap, about twice the expected value of
$0.41J$, was, by mistake, used for the single-particle dispersion in the
formulas of Ref. \cite{AffleckWeston}. As a result, the discrepancy between the
prediction of the nonlinear $\sigma$-model and the experimental results looks
larger than it actually is.

Figure 1 here is the corrected version of the Fig. 2 of Ref.
\onlinecite{Zaliznyak2001}, and is now in agreement with the Fig. 3 of that
paper.
Clearly, there is also much better agreement between the data and the
$\sigma$-model prediction now (note the absence of an arbitrary intensity
scaling).
Nevertheless, the observed continuum still lies noticeably {\it below} the
lowest energy of the two non-interacting purely 1D magnons, supporting our
previous conclusion that there is an {\it attraction} between the quasiparticles
which form the continuum states \cite{Zaliznyak2001}. Such attraction is
probably a result of the inter-chain coupling which leads to a 3D ordering in
CsNiCl$_3$. On the mean field level it can be understood in a manner similar to
the spinon attraction in the coupled S=1/2 chains, if the excitations are the
``spin-zero defects'' of Ref. \onlinecite{Gomez-Santos}, as shown in Fig. 2
below.

\begin{figure}[b] \vspace{-0.2in}
\parbox[b]{3.4in}{\psfig{file=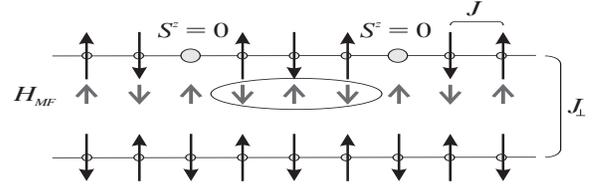,width=3in,height=1in}
\vspace{0.1in}\noindent%
\caption{ Attraction of the $S^z = 0$ defects in the
antiferromagnetic background, introduced by the mean field $H_{MF}$ arising from
the inter-chain ordering. }}
\end{figure} \vspace{-0.1in}

We thank Ian Affleck for pointing out this error, and F.~Essler for encouraging
discussions. This work was carried out under Contract DE-AC02-98CH10886,
Division of Materials Sciences, US DOE. The work on SPINS was supported by NSF
through DMR-9986442.

\end{document}